# Elastic and electronic tuning of magnetoresistance in MoTe$_2$

*Strain effects in MoTe$_2$*

Junjie Yang,[1] Jonathan Colen,[1] Jun Liu,[1] Manh Cuong Nguyen,[2] Gia-Wei Chern,[1] and Despina Louca[1,*]

[1]*Department of Physics, University of Virginia, Charlottesville, VA 22904, USA.*
[2]*Ames Laboratory, U. S. DOE, Iowa State University, Ames, IA 50011, USA.*

[*]*To whom correspondence should be addressed; E-mail: louca@virginia.edu*

## Abstract

Quasi-two dimensional transition metal dichalcogenides (TMD) exhibit dramatic properties that may transform electronic and photonic devices. We report on how the anomalously large magnetoresistance (MR) observed under high magnetic field in MoTe$_2$, a type II Weyl semimetal, can be reversibly controlled under tensile strain. The MR is enhanced by as much as ∼ 30 % at low temperatures and high magnetic fields, when uniaxial strain is applied along the *a*-crystallographic direction and reduced by about the same amount when strain is applied along the *b*-direction. We show that the large in-plane electric anisotropy is coupled with the structural transition from the 1T' monoclinic to the T$_d$ orthorhombic Weyl phase. A shift of the T$_d$ - 1T' phase boundary is achieved by minimal tensile strain. The sensitivity of the MR to tensile strain suggests the possibility of a nontrivial spin-orbital texture of the electron and hole pockets in the vicinity of Weyl points. Our ab initio calculations indeed show a significant orbital mixing on the Fermi surface, which is modified by the tensile strains.

## INTRODUCTION

Semiconducting TMDs exhibit many versatile physical characteristics and have become a new paradigm for optoelectronic applications (1-11), based on exfoliated single layer molecular structures (3-6). Fueled by intense interest on new device concepts, TMDs provide a platform from which optoelectronic properties such as spin-valley coupled physics and two-dimensional valley excitons can be explored (12). They have a desirable optical band gap in the 1-2 eV energy range, important for visible and near-infrared technologies. Manipulation of the band gap either by reducing the sample thickness down to a monolayer or applications of strain can lead to distinct changes in their physical characteristics. Demonstrated in $MoS_2$ (6), the reduction in the number of layers leads to a shift of the valance band maximum from the center of the Brillouin zone closer to the edge, reducing the band gap as it aligns with the conduction band. Excitons, electron-hole pairs, are strongly bound in TMDs with binding energies of the order of 0.6 eV because of an enhanced electron-hole interaction, and when crystal thickness is reduced down to a monolayer, exciton binding energy is lowered. Similarly, the gap can also be reduced under tensile strain leading to a transition to a metallic state.

Strain engineering is a widely used technique to control performance of electronic and spintronic devices (13), and has exposed a variety of strain-induced effects on the electronic behaviors including modulations of the band gap, carrier mobility and light absorption. Many proposed devices reported in the literature make use of the application and control of compressive or tensile strain. Examples include a piezoelectric substrate for the application of biaxial compressive strain (13), flexible substrates such as PET (polyethylene terephthalate) films on a three-point apparatus (14), or flexible PMMA (ePlastics) for uniaxial tensile strain applications (15), and bending of polycarbonate beams using a four-point apparatus (16). Most studies have focused on the photoluminescence properties between the 2H and 1T' phases. Little is known of the effects on the structural phase transition between 1T' and $T_d$ and the MR property. The observation of an extremely large MR in the layered TMDs of $WTe_2$ and $MoTe_2$ has led to a surge of interest in this field (4), but the effects on MR due to strain have not yet been reported.

The layered crystal structure consists of strong in-plane covalent bonding and weaker van der Waals type interactions between planes (17). $MoTe_2$ in particular (18) can exist in several crystal configurations that includes the hexagonal 2H, the metastable 1T' and the low temperature $T_d$ structures. $\beta$-$MoTe_2$ (1T') can be stabilized by quenching and is metallic with a monoclinic ($P2_1/m$) structure (17-19). Mo is surrounded by Te in an octahedral environment with Mo shifted off-center. It forms zigzag chains running along the b-axis that distort the Te sheets as shown in Fig. 1(A), resulting in a tilted c-axis with angle $\beta \sim 93^\circ$ (Fig. 1B). Upon cooling from room temperature, an anomaly appears in the transport data around 240 K that has been linked to a first order structural transition from the 1T' of $\beta$-MoTe$_2$ to the orthorhombic T$_d$ phase (Pmn2$_1$) (Fig. 1(B)). The resistivity anomaly can be seen in Fig. 1(C), measured along the b-crystallographic direction and is typical of what is found in the literature (18,19). The resistivity shows metallic behavior with a thermal hysteresis setting in below room temperature. Due to the in-plane anisotropy, the resistivity along the a-crystallographic direction is higher but shows the same essential feature (20). The T$_d$ phase exhibits the extreme MR effect and is the host of a Weyl semimetal state, a new state of matter in which collective excitations known as Weyl fermions may exist. Materials of this kind have unusual properties where linear dispersions of the valence and conduction bands cross at Weyl points (21,22). The band crossing is near the Fermi level forming a gapless node (23). It has been suggested that the band structure of MoTe$_2$ is highly sensitive even to small changes in the lattice constants (23) either brought upon by strain or as a function of temperature.

Strain modulated phase transitions have been studied in the TMDs especially in MoS$_2$. Using localized strain applied via wrinkling on atomically thin MoS$_2$ deposited on an elastomeric substrate, a reduction of the direct band gap was observed (24) presumably by shifting the conduction band. In MoTe$_2$, it was shown how the 2H-1T' phases can be reversibly switched by strain, coupled with a semiconducting to metal transition (6). By introducing variable tensile strain, $\varepsilon$, through the application of an electric field on a piezoelectric stack, we show that an anisotropy develops in the MR that sets in with the symmetry breaking of the 1T' and T$_d$ phases resulting in different responses depending on whether $\varepsilon$ is applied parallel to the a- or b-directions. To elucidate the effects on the MR, we report strain measurements of the in-plane transport anisotropy as a function of strain and magnetic field. The results provide a new venue in which the physical properties are tuned reversibly by strain without wrinkling the surface (6) or applying a contact force (13-16).

**RESULTS**

A schematic of the experimental setup used in our electric transport measurements on MoTe$_2$ is shown in Figs. 2(A) and (B). Measurements were carried out with the electric current fixed along the b-crystallographic axis and the magnetic field applied along c (25). Tensile strain was applied either along a- or b- as shown in Fig. 2(A) and 2(B), respectively. The temperature dependence of the resistivity as a function of applied field shows strikingly different responses when $\varepsilon$ // b versus $\varepsilon$ // a (Fig. 2(C) and (D)). At $\varepsilon$=0, the magnetic field brings about a significant increase in the resistivity below 50 K, yielding a very large MR in $MoTe_2$ as previously reported (1-8) (black line). The resulting $MR \equiv [\rho_{xx}(H,\epsilon)-\rho_{xx}(H=0,\epsilon=0)]/\rho_{xx}(H=0,\epsilon=0)$ reaches 38,539 % at 2 K and 9 T as shown in the inset. When $\epsilon // b$, a clear drop in the resistivity is observed, further decreasing under magnetic field, indicating that tensile strain along b suppresses MR. Interestingly, when $\epsilon // a$, the effect is reversed: the resistivity and consequently the MR are enhanced (Fig. 2(D)). The induced strain by the application of voltage is rather small, less than 0.05 % at the maximum. This behavior is rather typical of the samples that we measured. The two shown here are representative of the strain effects. However, when the samples are very thin, different behavior was sometimes observed due to the resistance dependence on $\epsilon$ at zero field. These results are reported in the Supplementary.

To further characterize the changes in MR induced by $\epsilon$, the MR was measured as a function of magnetic and tensile fields. The magnetic field dependence of the strain-induced change of MR (SMR), defined as $SMR=MR(\epsilon)-MR(\epsilon=0)$ is shown in Fig. 3. Interestingly, the SMR follows a near $(\mu_o H)^2$ behavior at all temperatures. It is negative when $\epsilon // b$ and positive when $\epsilon // a$. Consistent with the results shown in Fig. 2, strain along the a-direction enhances the MR, thus SMR is positive. Similarly, when strain is applied along the b-direction, the MR is reduced, hence the SMR is negative. Moreover, with cooling the strain effects become more pronounced as can be seen from the temperature dependence of the data shown in Figs. 3(A) and 3(B). In Figs. 3(C) and 3(D), the SMR at 9 T is plotted as a function of $\epsilon$ for $\epsilon // b$ and $\epsilon // a$, respectively. All curves exhibit a near linear behavior, which is expected from the near linear converse piezoelectric response of the piezoelectric stack used to apply the tensile strain. The negative and positive SMR behavior is seen here as well.

In addition to the non-saturated extremely large MR, $MoTe_2$ exhibits a structural phase transition $(1T' \rightarrow T_d)$ near 240 K, evidenced by the kink in the resistivity as mentioned above. Strain has a direct effect on the structural transition as well. Figs. 1(C) and 1(D) are plots of the transport $\rho(T)/\rho(280 \text{ K})$ and the derivative of $\rho(T)$ near the phase transition as a function of $\epsilon // b$, respectively. Similar results were obtained with $\epsilon // a$. Thermal hysteresis is seen on cooling and warming cycles in both crystallographic directions as well as a shift in the structural transition temperature with the applied tensile field. The shift is best seen in the derivative plot, $d\rho(T)/dT$, of Fig. 1(D). We define the maximum of $d\rho(T)/dT$ as $T_{S1}$ on cooling and $T_{S2}$ on warming. $T_{S1}(\epsilon) - T_{S1}(\epsilon=0)$ $(T_{S2}(\epsilon) - T_{S2}(\epsilon=0))$ follows a near linear dependence on $\epsilon$ on cooling (warming), but shifts in opposite direction depending on the crystallographic direction the strain is applied on (Fig. 1(E)). Furthermore, the strain induced change of the width of the hysteresis loop is defined as $HW(\epsilon)$-

HW(0) = $(T_{S2}(\varepsilon)-T_{S1}(\varepsilon))$ - $(T_{S2}(\varepsilon=0)- T_{S1}(\varepsilon=0))$.  When ε // b, the width grows while it shrinks when ε // a as seen in Fig. 1(F).   This suggests that the coexistence region of the 1T' - $T_d$ phases is stretched in temperature when ε // b, and compressed when ε // a.  Therefore, upon cooling, the 1T' →$T_d$ transition shifts down in temperature when ε // b, and shifts up when ε // a.  Similarly, on warming, the $T_d$ →1T' transition also shifts down in temperature when ε // b, and shifts up when ε // a.

To understand how the $T_d$ phase of $MoTe_2$ responds to uniaxial strain, we performed ab initio calculations based on density functional theory (DFT) with spin-orbit coupling included (26,27). Starting with the lattice constants obtained from experiment, the lattice structure under uniaxial tensile strain is determined taking into account the anisotropic Poisson ratio γ of $MoTe_2$. Our DFT calculation finds $\gamma_{ab}$=0.19 and $\gamma_{ac}$=0.96 for a tensile strain along a axis, and $\gamma_{ba}$=0.31 and $\gamma_{bc}$=0.54 for a strain along b axis. While these values agree on average with previous ab initio calculations assuming isotropic elastic constants (28), we emphasize that similar to its electronic behaviors, $MoTe_2$ exhibits rather anisotropic elastic properties.

The band structures of the undeformed lattice and that under a 0.5% strain along a- and b-directions are shown in Fig. 4(A)-(C). Modifications of the band dispersions are observed around the Γ point and along the Y→Γ direction allowing for a unique kind of band overlap manipulation through tensile strain application. In the case where strain is applied along the a-axis, the band shifts give rise to a density of states (DOS) that is somewhat reduced around $E_F$. Conversely, if the strain is applied along b, the bands also shift up or down but in the opposite way as shown in the figure and the DOS appears to be enhanced at $E_F$. The opposite trends of DOS change under different strains are consistent with the observed enhanced and reduced SMR.

The $k_z$=0 Fermi surface sheets in the vicinity of the Weyl points are shown in Fig. 4(D) - (F) for the case of undeformed lattice, and that under uniaxial strains along a and b directions, respectively. The Fermi level has a more pronounced response to strain along b-direction. In particular, the small electron pocket close to the Weyl points disappears at the Fermi level at 0.5% strain in the calculation. Also shown along the contours is the orbital character (d versus p orbitals) of the corresponding Bloch states. Our DFT calculation reveals significant *d-p* orbital mixing at the Fermi level, indicating a nontrivial orbital pseudo-spin texture of the Fermi surface.

While the DFT calculation provides an overall picture of how the band structures and Fermi surfaces are affected by strain, the relative change of the band parameters (of the order of 0.1 to 1 %) seems too small to explain the large SMR observed in our experiments. For example, interpolating our calculation to the experimental value of 0.05 % strain gives rise to a variation of band curvature of similar order. For a parabolic band, this variation indicates a similar 0.05 % change in the electron effective mass, which is consistent with the observed less than 0.01 % change in resistivity induced by strain at zero magnetic field. However, this band curvature variation seems too small to account for the observed 30 % SMR with an applied 0.05 % strain. In DFT calculations, the changes in band parameters mainly result from strain-induced modification of orbital overlaps and hopping integrals. Below we argue that the underlying mechanism of the large SMR is likely related to that of the extremely large MR itself in $MoTe_2$, which involves enhanced electron scattering by the magnetic field.

The extremely large MR (XMR) phenomenon observed in $MoT_2$ and a closely related compound, $WTe_2$, has been conventionally attributed to the perfect balance between the electron and hole populations (5,29). Indeed, this near-perfect compensation is crucial to the observed non-saturating $H^2$ increase of MR within the framework of two-carrier model. However, the compensation effect itself does not explain the extraordinary magnitude of XMR. For example, it has been shown that reducing the sample thickness through exfoliation (30) significantly suppresses the XMR in $WTe_2$ while electron and hole remain perfectly compensated. Later high resolution ARPES and magneto-transport experiments (31,32) also showed that the electron and hole densities are slightly imbalanced in $WTe_2$.

The underlying mechanism of XMR is likely due to a significant field-induced enhancement of electron backscattering (33). One scenario is that such backscattering is prohibited by symmetry at zero field, and the lifting of this protection by the magnetic field leads to the XMR. Microscopically, this protection could result from the nontrivial orbital and spin texture of the electron and hole pockets in proximity to a type-II Weyl point. Recent observations of circular dichroism by ARPES confirm such spin-polarized Fermi surfaces in $WTe_2$ (34,35). Indeed, significant d-p orbital mixing, an important ingredient for nontrivial orbital angular momentum texture, is obtained from our DFT calculation for $MoTe_2$ (see Fig.4). More importantly, the electron and hole pockets inherit the chiral nature of the Weyl point and exhibit a definite helicity between their momentum and spin. Consequently, carrier scattering between pockets of opposite chiralities, especially those on the opposite sides of the $\Gamma$ point, is suppressed due to the opposite sign of Berry phases from time-reversal related scattering paths (36). In the presence of a magnetic field, the broken time-reversal symmetry leads to an imperfect cancellation of Berry phases and an enhanced inter-pocket backscattering (36).

As orbitals directly couple to the lattice, applying strain to a Weyl semimetal is expected to modify the orbital texture of electron and hole pockets, hence affecting the inter-pocket electron scattering. Our DFT calculation shows that applying strains indeed alters the orbital character of the electron and hole pockets (see Fig. 4(C)--(D)). Here we attribute the observed large SMR to the strain-induced modification of the pocket orbital texture, hence promoting or suppressing the inter-pocket backscattering. Indeed, recent theoretical studies have shown that elastic lattice deformations couple to the electronic degrees of freedom as pseudo-gauge fields in Weyl semimetals (37,38). Moreover, since the tilting of the Dirac cone is along the *a* direction in MoTe$_2$, we expect rather different effects from tensile strains applied along the a and b directions. Finally, we want to emphasize that the strain-induced pseudo-gauge fields preserves the time-reversal symmetry contrary to the real magnetic field. This explains the rather small strain-induced resistivity change at zero magnetic field. More precisely, the observed large SMR thus originates from a nontrivial interplay between the real and pseudo-gauge fields induced by magnetic field and strain, respectively. We hope our experiments will motivate further theoretical investigations into this novel phenomenon.

## Materials and Methods

Sample synthesis: High purity elements of Mo (99.9999%) and Te (99.9999 %) were weighted and placed in a quartz tube with a ratio of 1:25 for single crystal growth. The quartz tube was subsequently heated up to 1050 °C and held for 24 hours. Then, the quartz tube was slowly cooled down to 900 °C, followed by quenching into liquid nitrogen. Single crystals of MoTe$_2$ were grown from the Te flux.

Measurements: Single crystal MoTe$_2$ was attached on the surface of a piezoelectric stack using ultra high-strength 2- component epoxy glue (UHU, Germany). The epoxy glue was subsequently cured at 80 °C by 1 hour. The sample and piezo stack are well electrically isolated since the epoxy glue is a very good insulator. Gold wires were then attached on the surface of MoTe$_2$ single crystal using silver epoxy. Silver epoxy was cured at 80 °C for 3 hours before the measurements. Electric transport measurements were performed in a Quantum Design Physical Properties Measurement System (PPMS) in a temperature range from 2 to 300 K. Strain was measured using a strain gauge which was glued on the other side of the piezoelectric stack.

Calculations: We use Density Functional Theory implemented in the Vienna Ab-initio Simulation Package (VASP) for a material specific calculation to study the uniaxial tensile strain effect of the T$_d$ phase of MoTe$_2$. VASP is a plane-wave based projector-augmented wave (PAW) pseudo-potential method with its exchange-correlation functional taking the generalized-gradient approximation (GGA) parameterized by Perdew-Burke-Ernzerhof (PBE). Specifically,

ENCUT=300eV, K-points=12*6*3 generated with automatic K mesh, and NBANDS=128 with spin-orbital coupling switched on. The convergence in basis set and k-mesh size is verified. To simulate the effect of unaxial tensile strain of this material, we start with the lattice constant obtained from experiment, a = 3.477 Å, b = 6.335 Å, c = 13.883 Å, and take its elastic property under full consideration. We use VASP to calculate its anisotropic Poisson ratios, $\gamma$, for uniaxial tensile strains, and then apply these Poisson ratios to determine the corresponding lattice constants under 0.5% tensile strain along a and b axes. Specifically, our DFT calculation finds $\gamma_{ab}$=0.19 and $\gamma_{ac}$=0.96 along a-axis, and $\gamma_{ba}$=0.31 and $\gamma_{bc}$=0.54 along b-axis. The electronic structure calculation is carried out on these more realistic lattice structures, and the evolution of the band structure and density of state (DOS) under tensile strain is revealed. The Weyl points are checked to be consistent with previous publications and are indicated in Fig 4 with filled dots. Wannier90 is used to fit the effective KS Hamiltonian on a very dense k space mesh, and a home-made program maps out the Fermi surface morphologies of the relevant structures at $k_z = 0$. These Fermi surface K points are compiled into a KPOINTS file for a second pass of VASP which finally produces the mixed local p-d orbital ratio characters of the electron and hole pockets presented in the paper.

## Acknowledgments

**Funding:** This work has been supported by the Department of Energy, Grant number DE-FG02-01ER45927.

**Author contributions:** JY and DL devised the experiment. JY and JC did the sample preparation and JL, MCN and GWC did the calculations. DL wrote the paper with contributions from GWC.

**Competing interests:** The authors declare that they have no competing interests.

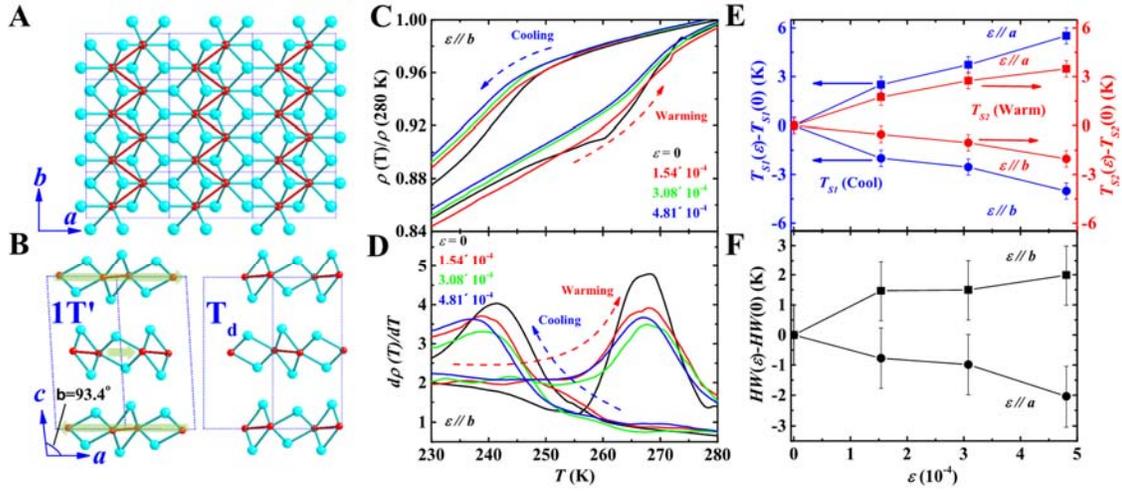

**Fig. 1. Structure and transport.** (**A**) The crystal structure of MoTe$_2$ in the 1T' phase projected on the ab-plane with the zig-zag chains marked running along the b-axis. (**B**) The unit cell of the 1T' and T$_d$ structures projected on the ac-plane. (**C**) A plot of the resistivity, ρ(T), along the b-axis with ε∥b. The kink around 240 K is evidence of the structural phase transition from the 1T' to the T$_d$ phase. (**D**) The derivative of dρ(T)/dT calculated from (**C**). (**E**) The strain dependence of T$_{S1}$(ε) - T$_{S1}$(ε=0) on cooling (left axis) and T$_{S2}$(ε) - T$_{S2}$(ε=0) on warming (right axis). T$_{S1}$/T$_{S2}$ are the maximum obtained from the dρ(T)/dT derivative on cooling and warming cycles. (**F**) The strain dependence of the width of the thermal hysteresis, defined as HW (ε) – HW (0) = (T$_{S2}$(ε)-T$_{S1}$(ε)) - (T$_{S2}$(ε=0) - T$_{S1}$(ε=0)), with ε applied along a- and b-directions.

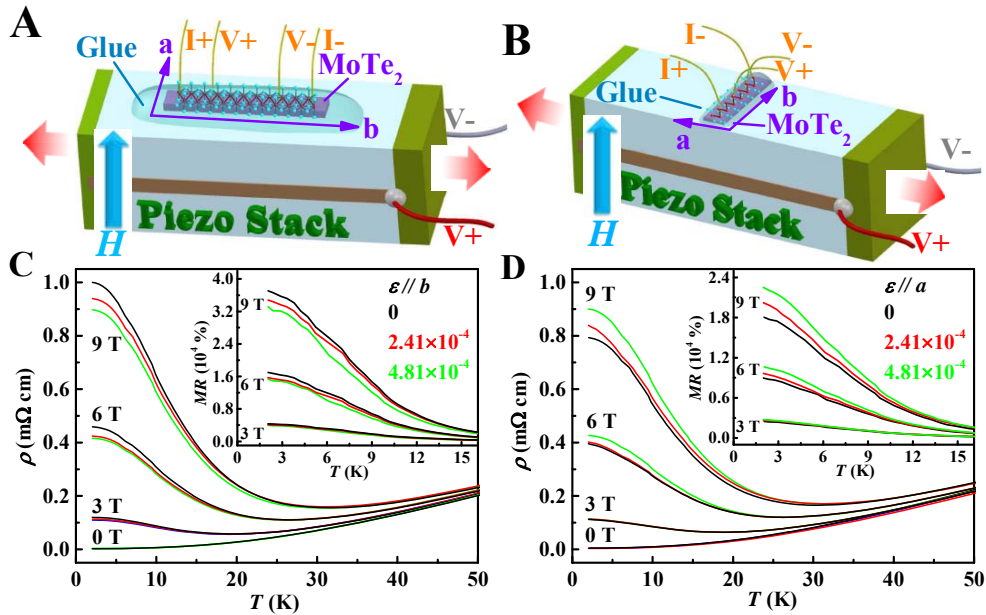

**Fig. 2. Magnetoresistance under field.** (**A**) and (**B**) Schematic illustrations of the electric transport measurements with strain along b and a crystallographic directions, respectively. The red arrows indicate the expansion directions for the piezo-stack under electric voltage. A ribbon-like MoTe₂ single crystal was glued to the surface of a piezo-stack and cured so that it can transfer the strain effectively. Four gold wires were attached to the surface of the crystal for the four probe electric transport measurements. Tensile strain was applied on MoTe₂ through a converse piezoelectric effect which can be controlled by applying electric field on the piezo-stack. (C) and (D) Plots of the in-plane resistivity at 3, 6 and 9 T as a function of strain.

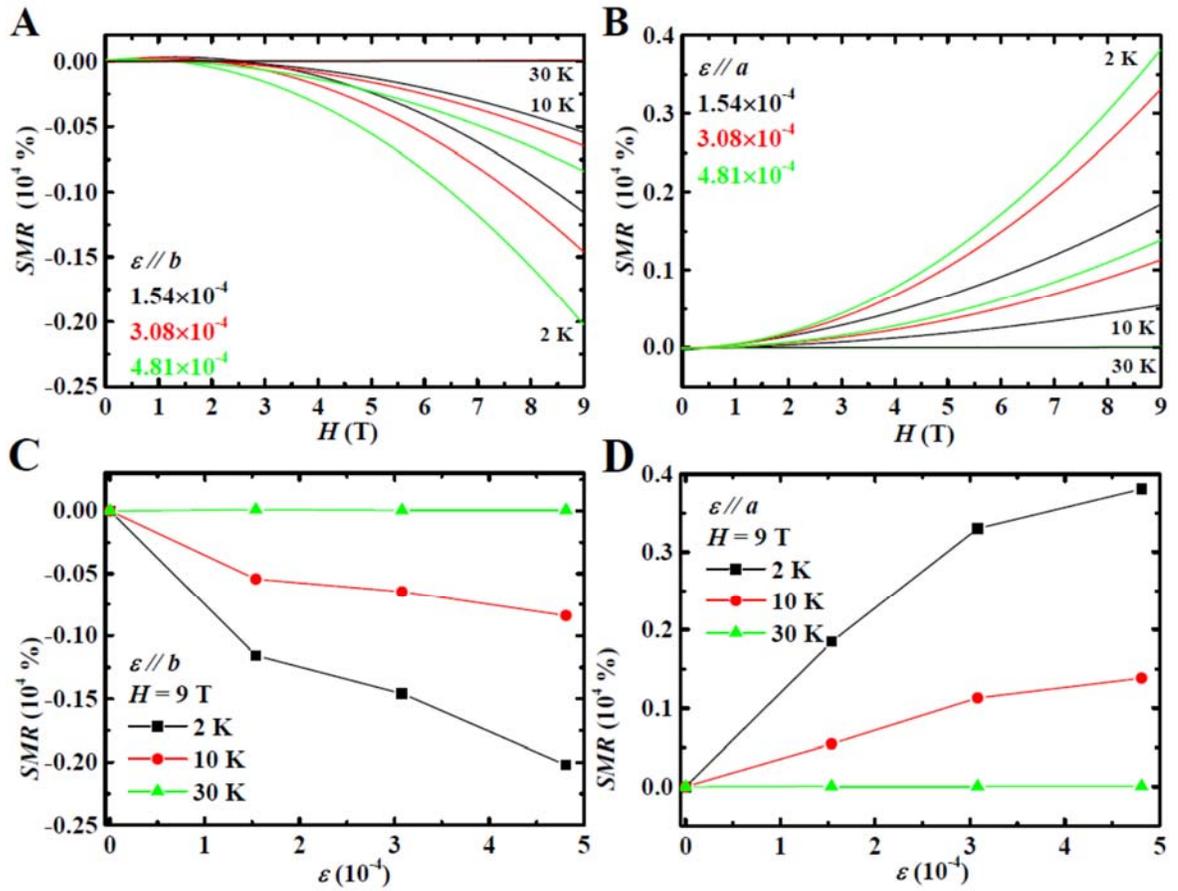

**Fig. 3. Strain induced MR.** (**A**) The magnetic field dependence of SMR determined at several temperatures and as a function of ε ∥ b. The SMR is negative as the MR is reduced under strain. In (**B**), the magnetic field dependence of SMR is determined with ε∥a. In (**C**) and (**D**) The SMR is plotted as a function of tensile strain, ε, at 9 T at three different temperatures.

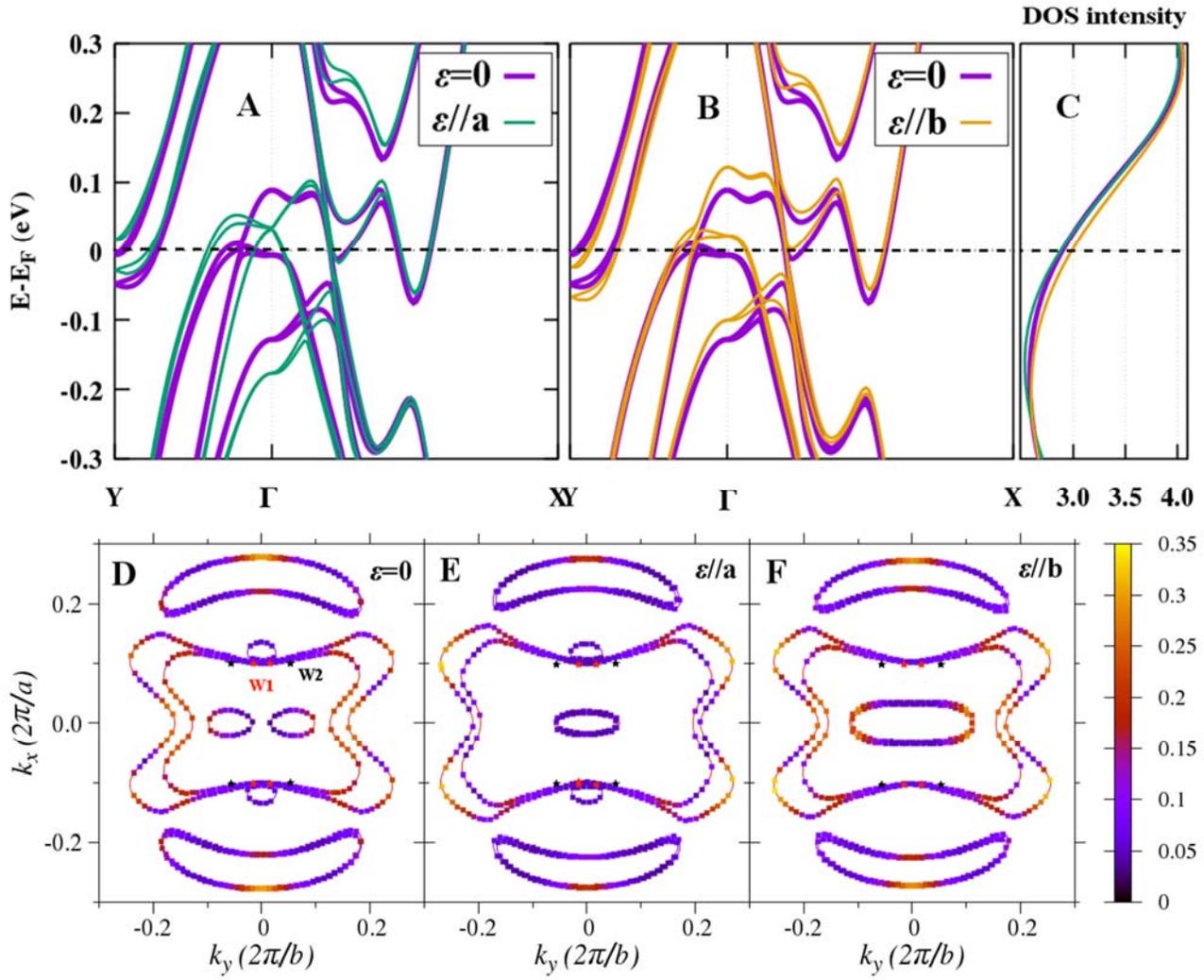

**Fig. 4. Band structure under strain.** Band structure comparison of the $T_d$ phase of MoTe$_2$ for strains along a and b directions are shown in (**A**) and (**B**), respectively. The corresponding changes of the DOS are shown in panel (**C**). The Fermi surface cuts at $k_z = 0$ are shown in (**D**), (**E**), and (**F**) for zero strain and 0.5% strains along a and b directions, respectively. The two types of Weyl points are indicated with red and black dots between the electron and hole pockets near the zone center in (**D**). The color gradient indicates the component ratio of the Mo d-orbitals to Te p-orbitals around the Fermi surfaces.

# Supporting material online for

## Elastic and electronic tuning of magnetoresistance in MoTe$_2$

Junjie Yang, Jonathan Colen, Jun Liu, Manh Cuong Nguyen, Gia-Wei Chern, Despina Louca

**This PDF file includes :**

# Supporting Online Material

**XRD for MoTe₂ single crystals.**

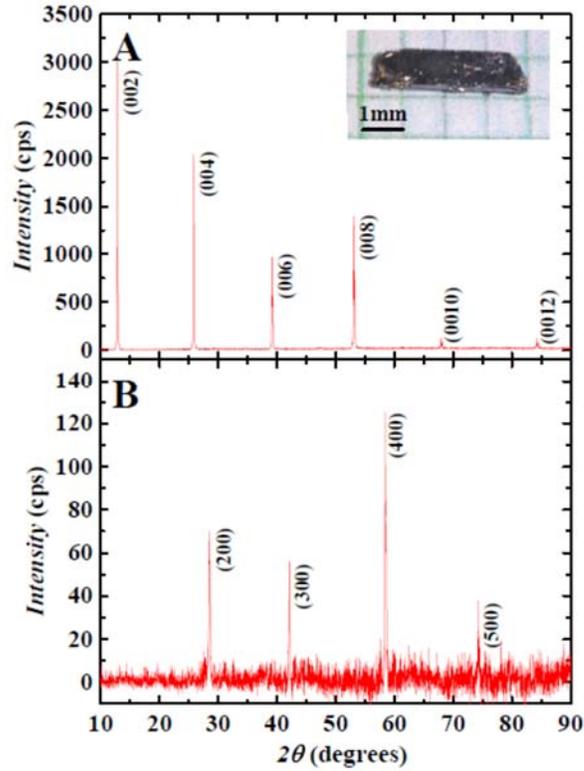

**Figure S1.** XRD of MoTe₂ single crystals aligned along: (a) (00L) and (b) (H00). The inset in (a) is a picture of a MoTe₂ single crystal.

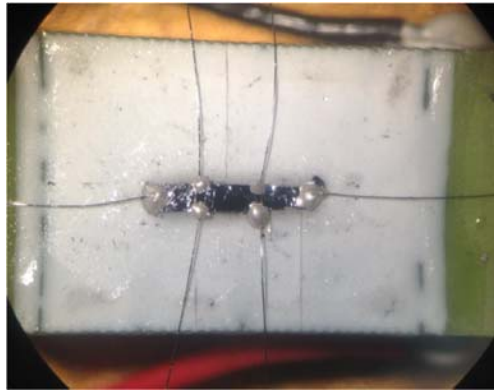

**Figure S2.** MoTe₂ crystal on the piezoelectric stack.

*SMR* **of different samples.**

Figure S3 exhibits the electrical transport data under different strain for two different samples. All results were measured at 2 K, and the strain was applied along the $b$ direction. *SMR* is defined as $SMR = \frac{[R(\varepsilon,H)-R(\varepsilon,0)]-[R(0,H)-R(0,0)]}{R(0,0)} \times 100\%$. Fig. S3 A and B show resistance vs magnetic field curves. Fig. S3 C and D show the calculated SMR as a function of magnetic field for the two samples. As it can be seen from Fig. S3 C and D, the strain effect mainly appears at high magnetic fields and varies with sample. Sample 2 exhibits a much bigger strain effect than sample 1. Fig. S4 are plots of the electrical transport data under different strain with strain along *a* for two different samples. The strain effect mainly appears at high magnetic field and it varies with sample. Sample 4 exhibits a much bigger strain effect than sample 3.

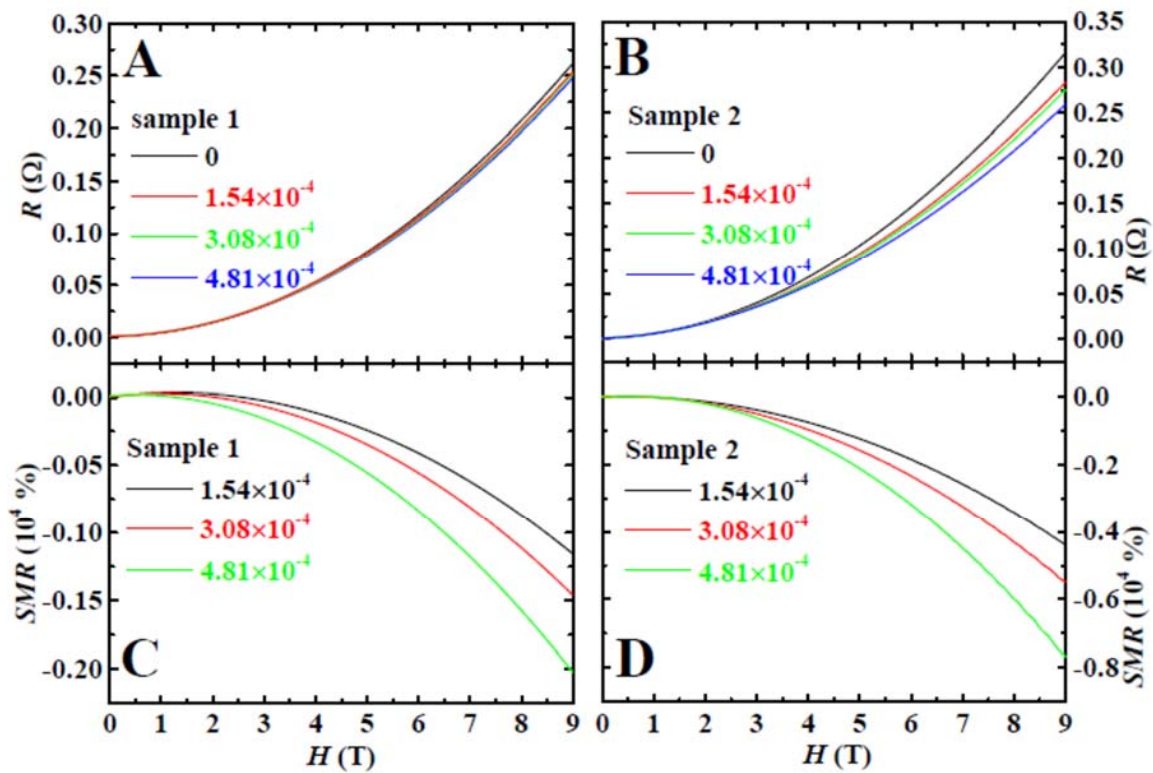

**Figure S3.** Electrical transport data measured at 2 K for two different samples: (A), *R-H* curves under different strain for sample 1, (B) *R-H* curves under different strain for sample 2, (C) *SMR* as a function of magnetic field for sample 1 and (D) *SMR* as a function of magnetic field for sample 2. Strain was applied along *b* direction.

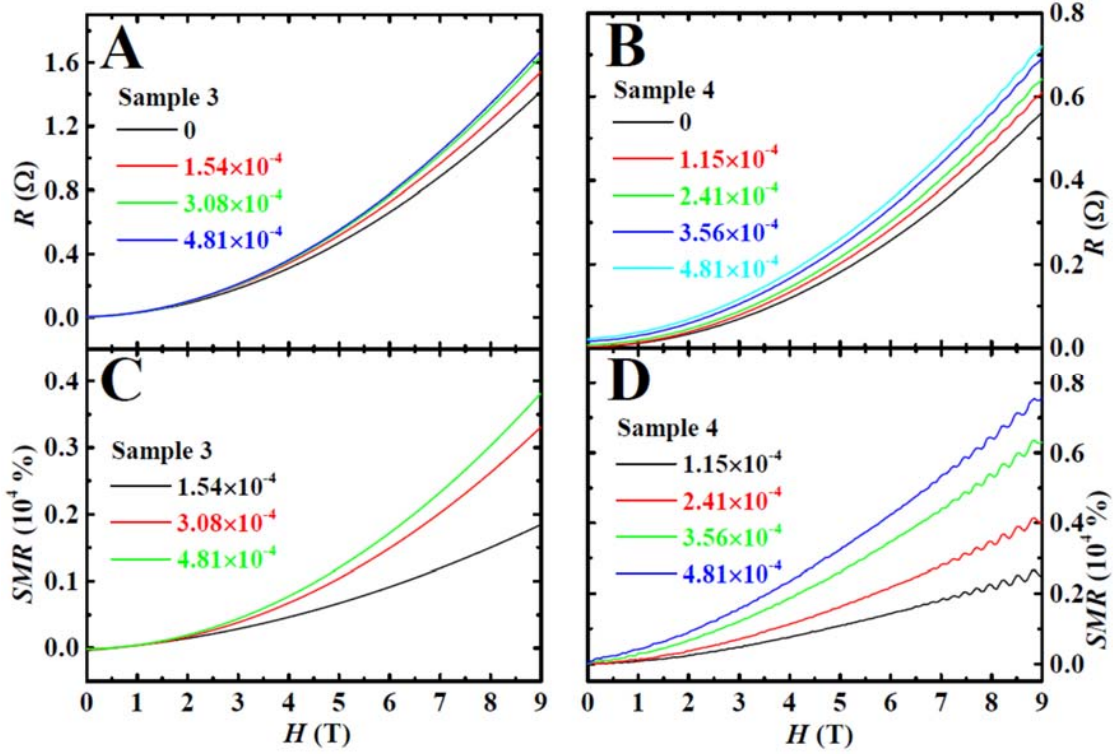

**Figure S4.** Electrical transport data measured at 2 K for two different samples: (A), *R-H* curves under different strain for sample 3, (B) *R-H* curves under different strain for sample 4, (C) *SMR* as a function of magnetic field for sample 3 and (D) *SMR* as a function of magnetic field for sample 4. Strain wass applied along *a* direction.

### Hall Effect and Two-band model fitting

Figure S5 shows the results of Hall effect and two-band model fitting. The electrical transport data was measured at 2 K with strain applied along *b* direction. Fig. S5 A and B show the Hall conductivity $\sigma_{xy} = -\rho_{xy}/(\rho^2_{xy} + \rho^2_{xx})$ and the longitudinal conductivity $\sigma_{xx} = \rho_{xx}/(\rho^2_{xy} + \rho^2_{xx})$ measured without strain. According to the two-band model (29, 39-41), the Hall conductivity is $\sigma_{xy} = \left[ \frac{n_e \mu_e^2}{1+(\mu_e \mu_0 H)^2} - \frac{n_h \mu_h^2}{1+(\mu_h \mu_0 H)^2} \right] e \mu_0 H$, and the longitudinal conductivity is $\sigma_{xx} = \frac{n_e e \mu_e}{1+(\mu_e \mu_0 H)^2} + \frac{n_h e \mu_h}{1+(\mu_h \mu_0 H)^2}$. The red lines shown in Fig. S5 A and B are the fitting curves. Fig. S5 C-F show the extracted carrier density, n, and mobility, $\mu_e$ for electrons and $\mu_h$ for holes. Both carrier density and mobility decrease with increasing strain when strain is applied along the *b* direction.

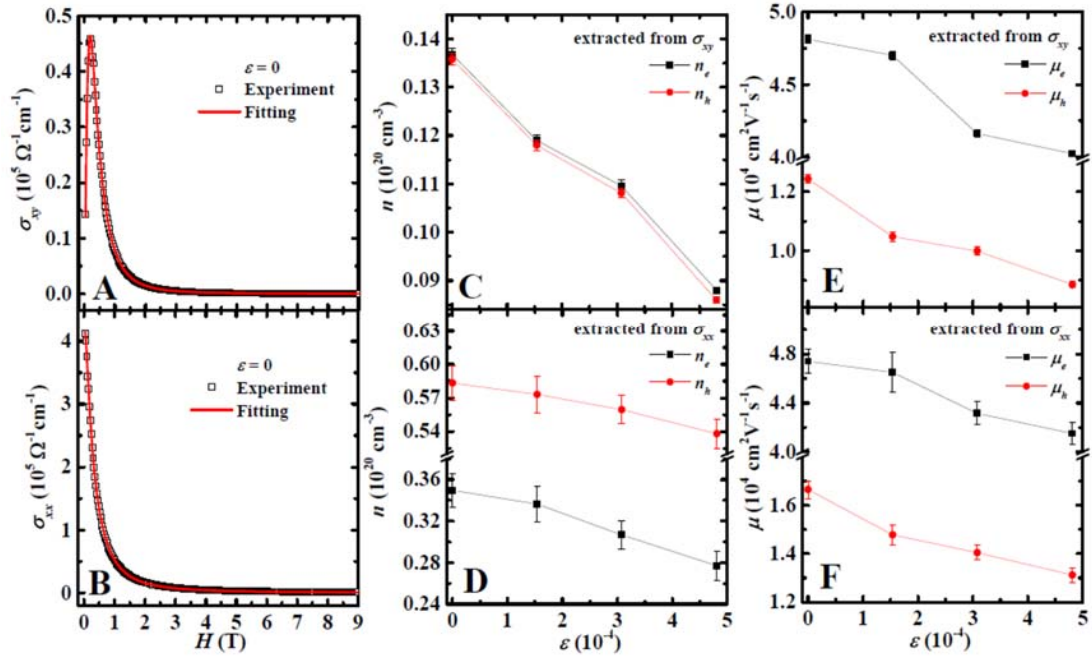

Figure S5. (A) Hall conductivity $\sigma_{xy}$ and (B) the longitudinal conductivity $\sigma_{xx}$ measured without strain at 2 K. (C) electron and holes density extracted from the fitting of $\sigma_{xy}$, (D) electron and holes density extracted from the fitting of $\sigma_{xx}$, (E) electron and holes mobility extracted from the fitting of $\sigma_{xy}$ and (F) electron and holes mobility extracted from the fitting of $\sigma_{xx}$. Strain was applied along $b$ direction.

Strain vs Voltage

Table S1. The piezoelectric stack shows an almost linear behavior in the voltage range from 0 V to 150 V (25). Hence, we used the maximum response of the piezoelectric stack at 150 V to estimate the strain values for different voltages: $\varepsilon = \varepsilon_{max} \times V / 150$. $\varepsilon_{max} = 14.44 \times 10^{-4}$.

| Voltage (V) | 12 | 16 | 25 | 32 | 37 | 50 |
|---|---|---|---|---|---|---|
| Strain ($10^{-4}$) | 1.15 | 1.54 | 2.41 | 3.08 | 3.56 | 4.81 |